# Thermo-mechanical behaviour of a compacted swelling clay

Anh-Minh Tang[1], Yu-Jun Cui[1], and Nathalie Barnel[2]

[1] Researcher and Professor respectively, École Nationale des Ponts et Chaussées, Paris (CERMES – Institut Navier).

[2] Researcher, Électricité De France.

**Corresponding author**

Prof. Yu-Jun Cui
ENPC/CERMES
6 et 8, av. Blaise Pascal
Cité Descartes, Champs-sur-Marne
77455 Marne La Vallée cedex 2
France
Tel: 33 1 64 15 35 50
Fax : 33 1 64 15 35 62
E-mail : cui@cermes.enpc.fr

**Abstract**

Compacted unsaturated swelling clay is often considered as a possible buffer material for deep nuclear waste disposal. An isotropic cell permitting simultaneous control of suction, temperature and pressure was used to study the thermo-mechanical behaviour of this clay. Tests were performed at total suctions ranging from 9 to 110 MPa, temperature from 25 to 80 °C, isotropic pressure from 0.1 to 60 MPa. It was observed that heating at constant suction and pressure induces either swelling or contraction. The results from compression tests at constant suction and temperature evidenced that at lower suction, the yield pressure was lower, the elastic compressibility parameter and the plastic compressibility parameter were higher. On the other hand, at a similar suction, the yield pressure was slightly influenced by the temperature; and the compressibility parameters were insensitive to temperature changes. The thermal hardening phenomenon was equally evidenced by following a thermo-mechanical path of loading–heating-cooling-reloading.

**Introduction**

Heavily compacted bentonite is often proposed to be used for engineered barriers in high-level nuclear waste disposals. In this condition, this swelling clay, which is initially in an unsaturated state, is subjected to thermo-hydro-mechanical actions, for exemple: (1) heat dissipation from the nuclear waste packages; (2) water infiltration from the geological barrier; (3) stresses generated by the swelling of the engineered barrier in confined conditions. A deep understanding of the behaviour of this clay under these coupled thermo-hydro-mechanical actions is essential to make a safe conception of the whole storage system.

The hydro-mechanical behaviour of compacted swelling soils at ambient temperature have been studied in previous works (Al-Mukhtar *et al.* 1999; Belanteur *et al.* 1997; Bucher & Mayor 1989; Cui *et al.* 2002; Cuisinier & Masrouri 2004; Delage *et al.* 1998; Komine & Ogata 1994; Lloret *et al.* 2003; Villar 1999; etc.). A common conclusion from these works is that the soil mechanical properties are influenced by the soil suction: the higher the suction, the higher is the yield pressure ($p_0$). In addition, the effect of suction on the elastic compressibility parameter ($\kappa$) and the plastic compressibility parameter ($\lambda(s)$) (the notations adopted for theses parameters are the same as in the Barcelona Basic Model elaborated by Alonso et al. 1990), determined from oedometer tests, was also observed. Regarding the swelling properties, it was evidenced that swelling is produced during wetting under low stresses and on the contrary collapse is induced under high stresses. When wetting under a confined condition (constant volume), swelling pressure is generally developed. These swelling properties are related to the presence of highly active clay minerals. During wetting, water molecules enter between the clay layers, inducing the swelling of clay aggregates and therefore a macroscopic swelling response.

As far as the thermo-mechanical behaviour is concerned, most works have been done on saturated clays (Akagi & Komiya 1995; Baldi *et al.* 1988; Burghignoli *et al.* 2000; Campanella & Mitchell 1968; Cekerevac & Laloui 2004; De Bruyn & Thimus 1996; Del Olmo *et al.* 1996; Delage *et al.* 2000; Demars & Charles 1982; Eriksson 1989; Habibagahi 1977; Hueckel & Baldi 1990; Kuntiwattanakul *et al.* 1995; Laloui 2001; Noble & Demirel 1969; Shimizu 2003; Sultan *et al.* 2002; Tidfors & Sallfors 1989; Towhata *et al.* 1993; etc.). It has been observed that heating may induce expansion under low stresses (or large over-consolidation ratio *OCR*) and contraction under high stress (or small *OCR*). The effects of temperature on the mechanical properties observed are often contradictory. Nevertheless, it can be concluded that the temperature effect on mechanical properties is quite small. Cui *et al.* (2000) described two phenomena produced during heating: (1) expansion of soil constituents (solid and water); (2) mechanical weakening of the contacts between soil aggregates. The expansion of the soil components explains the phenomenon of macroscopic thermal expansion under low stresses, as mentioned before; the mechanical weakening of contacts explains the thermal contraction under high stresses. This thermal volume change phenomenon has a significant effect on the soil shear strength. On one hand, the expansion of aggregates induces a decrease of soil strength, but on the other hand, the thermal contraction hardens the soil and increases the shear strength. These two opposed phenomena explain the conflicting results in the literature in terms of temperature effect on shear strength.

Regarding the thermo-mechanical behaviour of unsaturated soil, Villar & Lloret (2004) studied the temperature effect on the hydro-mechanical behaviour of a compacted expansive clay (FEBEX bentonite, liquid limit $w_L = 102\%$, plasticity index *PI* = 52%) in an oedometer. During wetting, it was observed that at higher temperatures, the swelling strain under constant pressure reduced, as did the swelling pressure under a constant volume condition. The same phenomena have been observed by Romero *et al.* (2003) on Boom clay ($w_L = 56\%$, *PI* = 27%), also in an oedometer. Saix *et al.* (2000) performed compression tests in a triaxial cell at constant suction (4.9 kPa) and temperature (30 – 70°C) on a clayey silty sand ($w_L = 25\%$, *PI* = 10.5%). A decrease followed by an increase of $p_0$ with increasing temperature was observed, the minimum value corresponding to a temperature of 45 °C. With compacted FEBEX bentonite, Romero *et al.* (2005) observed that the stiffness upon loading at 14 MPa suction increased with temperature. On the contrary, with compacted Boom clay, Romero *et al.* (2003) noticed that the compression index is larger at higher temperatures, but the swelling index seemed to be temperature independent. In terms of thermal volume changes, Saix *et al.* (2000) observed a contraction during heating of the clayey silty sand under constant stress at 42, 160 and 800 kPa in oedometer. On compacted FEBEX bentonite and Boom clay, Romero *et al.* (2005) observed a thermal expansion under low stresses.

In order to understand better the thermo-mechanical behaviour of unsaturated soils, in the present work, unsaturated compacted MX80 bentonite was studied using a suction-temperature controlled isotropic cell. The effect of suction and temperature on soil compressibility was studied by carrying out compression tests; the effect of suction and stress on the thermal volume change behaviour was investigated by performing heating/cooling tests.

**Material**

The MX80 bentonite is a clay from Wyoming (USA). With its high content of montmorillonite (80%), it has a liquid limit $w_L = 520\%$, a plastic limit $w_P = 46\%$, and can be classified as a highly expansive clay. Its specific gravity is $G_s = 2.76$. The cation exchange capacity (*CEC*) is 76 meq/100g (83% Na, 11% Ca, 5% Mg, 1% K).

Prior to the compaction, the clay was sieved at 2 mm and air-dried at 44% relative humidity (*RH*) and 20 °C temperature (*T*). That corresponds to a total suction of 110 MPa (see Tang & Cui 2005). At equilibrium, its water content $w = 10\pm2\%$. The clay was first compacted in a neoprene tube (90 mm in diameter, 120 mm in length, 1.2 mm in thickness) with a closed end. The sample within the tube was then compressed in an isotropic cell under a static pressure of 40 MPa. After this compression, clay specimens were placed back in the chamber at $RH = 44\%$ and $T = 20$ °C. This procedure allowed the compacted specimens to have a dry density $\rho_d = 1.78\pm0.3$ Mg/m$^3$, a void ratio $e = 0.55\pm0.03$, and a degree of saturation $S_r = 50\pm1\%$.

**Methods**

A suction-temperature controlled isotropic cell has been developed (Tang et al. 2006). The basic scheme of the cell is presented in Figure 1. The soil specimen (80 mm in diameter, 10 – 15 mm in height) is sandwiched between two dry porous stones, which are confined within two metallic plates. Several holes (2 mm in diameter, 7 mm in spacing) are drilled in the lower plate, allowing the moisture exchange between soil specimen and the chamber below the lower plate. Within this chamber, there is a glass cup containing an over-saturated saline solution to control the relative humidity in the chamber and therefore that in the soil sample. Measurements performed by Tang & Cui (2005) in a chamber having similar dimensions showed that the relative humidity in the chamber reached the target value in less than 5 hours. Measurements of suction generated using over-saturated salt solutions at different temperatures in this other chamber were performed by Tang & Cui (2005) and the values were applied in the present work to determine indirectly the imposed suction values. The soil specimen and the two metallic plates are covered by a neoprene membrane (1.2 mm thickness) that avoids any exchange between the confining water in the cell and the soil pore water. A thermocouple installed inside the cell is used to monitor of the ambient temperature. Several O-rings are used in different places for waterproof purpose.

The experimental setup is presented in Figure 2. The cell is immersed in a temperature controlled bath ($T = 25 - 90$ °C). The temperature probe installed in the cell is connected to the data logger. A volume/pressure controller controls the confining pressure in the cell. This controller also monitors the soil volume change through the volume change of confining water in the cell. The other temperature-controlled bath ($T = 25$ °C) avoids the entry of hot water from the cell to the volume/pressure controller. During each test, the volume and pressure of water in this controller are recorded by the data logger. The room temperature is controlled at 20±1 °C.

Prior to any test, compacted soil specimens (having an initial total suction of 110 MPa) were machined to obtain the required dimensions (80 mm in diameter, 10 – 15 mm high). For the tests performed at suctions lower than 110 MPa, the soil specimens were first wetted using the vapour equilibrium technique (see Delage *et al.* 1998) in a sealed box in order to reduce the suction to the desired values. The dimensions of the specimen at equilibrium after wetting were measured using a precision calliper, allowing the radial and axial strains upon wetting to be determined. After the initial wetting, soil specimens are installed in the cell for thermo-mechanical tests. Before the installation, the samples were machined again to fit the required diameter (80 mm). During the thermo-mechanical tests, the temperature was varied from 25 °C to 80 °C and pressure from 0.1 to 60 MPa, and three suctions of 9, 39, and 110 MPa were considered. Eleven tests were performed from the same initial conditions ($p = 0.1$ MPa, $\psi$ = 110 MPa, $T$ = 25 °C), and the start and end points stress paths of these tests are presented in Table 1.

During a compression test, the confining pressure of the cell is increased step by step: $p$ = 0.1, 0.2, 0.5, 1, 2, 5, 10, 20, 50 MPa, and the water volume change of the volume/pressure controller is recorded. An example of the results obtained from the step 1 to 2 MPa is presented in Figure 3 with water volume change plotted versus logarithm of time. It can be observed that the volume decreased rapidly at the beginning and stabilised after 3 days. The consolidation is considered to be stabilised when the volume change during 24 h is less than 50 mm$^3$ (corresponding to a soil volumetric strain of 0.1%). The water volume change during a loading step corresponds to the sum of the soil volume change and the expansion of the tubing system and the cell due to pressure increase. The calibration test was performed with a metallic specimen that had the same dimensions of the soil specimen. It can be seen that the calibration curve rapidly stabilised, in less than 10 minutes. In fact, as the calibration test is performed on a metallic specimen, the response is almost instantaneous, while in the test with soil specimen, 3 days was needed to reach the stabilisation in this example. The volume change of the soil specimen ($dV$) can be determined as the difference between the two curves, $dV$ = -1200 mm$^3$ in this case.

The determination of the soil volume change during heating is presented in Figure 4. During heating from 25 °C to 70 °C under a constant pressure of 0.1 MPa, water in the cell expanded and moved into the volume/pressure controller. The volume of water inside the controller increased from this flow by approximately 11 000 mm$^3$. Assuming that the coefficient of thermal expansion of the metal sample ($10^{-6}$ °C$^{-1}$) is negligible compared with that of the compacted clay soil (about $10^{-4}$ °C$^{-1}$, after Romero *et al.* 2005), the soil volume change can be determined from the difference between the results obtained from the test and the calibration. In this case, during heating from 25 °C to 70 °C, the soil volume increased by 500 mm³, which corresponds to a volumetric strain, $\varepsilon_v$ = -0.9 % (in this example, the soil specimen dimensions were: 80 mm in diameter and 11 mm in height).

As far as the suction control is considered, the soil suction is assumed to be equal to the total suction imposed by the salt solution when no volume change is observed at constant pressure and temperature. Indeed, as the soil studied (MX80 bentonite) is a highly expansive clay, suction changes usually induce volume changes. As mentioned by Tang & Cui (2005), temperature change induces a change in the suction imposed by a saturated saline solution. However, in the test with the isotropic cell, thermal loadings are usually applied during a short duration (less than 24h), within this short duration the suction in the soil is assumed to remain unchanged during thermal loading. According to Tang & Cui (2006), soil specimen having similar dimensions needed more than two weeks to reach the suction equilibrium after a change of suction imposed by salt solution. In addition, when the over-saturated NaCl solution is used, the suction is considered to be constant with temperature change. Three suctions were considered: $\psi$ = 9 MPa (KNO$_3$ at 25 °C and K$_2$SO$_4$ at 80 °C); 39 MPa (NaCl at 25, 40, and 60 °C); and 110 MPa (K$_2$CO$_3$ at 25 °C, MgNO$_3$ at 80 °C). The salt solution installed in the cell is chosen as a function of the desired values of suction and temperature during the mechanical loading path.

**Results**

Volume change under wetting

The volumetric strain of the soil specimens during wetting from the initial suction (110 MPa) to 39 and 9 MPa is presented in Figure 5 together with axial and radial strains. The volume of the soil specimen increased 50% when the suction was decreased to 9 MPa. In addition, it can be observed that the radial strain was equal to the axial strain, showing an isotropic behaviour. This observation is consistent with the isotropic compaction procedure adopted for the specimen preparation.

Volume change under thermal loading

Figure 6 shows the stress paths followed by tests T1, T2, T3, T4 and T5, in a space of total suction ($\psi$), pressure ($p$), and temperature ($T$). The initial state is defined by a low confining pressure $p = 0.1$ MPa, the room temperature $T = 25$ °C and a total suction of 110 MPa. For test T1, the soil specimen was heated to 60 °C under $p = 0.1$ MPa. For test T2, a decrease of suction from 110 to 39 MPa at $T = 25$ °C was first undertaken before heating to $T = 70$ °C. For test T3, the soil specimen was wetted to $\psi = 9$ MPa and then underwent a thermal cycle $T = 25 - 80 - 25 - 80$ °C. For test T4, the pressure was first increased to $p = 5$ MPa; a thermal cycle $T = 25 - 80 - 25$ °C was then applied while $p$ was kept constant. For test T5, a wetting to $s = 39$ MPa was first applied, followed by a loading to $p = 5$ MPa. Afterwards, a thermal cycle $T = 25 - 80 - 25$ °C was undertaken under constant pressure (5 MPa).

The results on thermal volume change under $p = 0.1$ MPa are presented in Figure 7. The results from the tests T1 ($\psi = 110$ MPa) and T2 ($\psi = 39$ MPa) show that heating induced an expansion. Considering that this expansion is linear, a coefficient of thermal expansion $\alpha = 2 \times 10^{-4}$ °C$^{-1}$ can be deduced. The result from test T3 ($\psi = 9$ MPa) shows that, on the contrary, heating induced a volume decrease. In addition, the subsequent cooling – reheating cycle undertaken shows a reversible behaviour. Note that significant data scatter was observed in this test and it is difficult to quantify the volume change behaviour during the cooling-reheating stage.

Figure 8 presents the results obtained from the tests T4 ($\psi$ = 110 MPa) and T5 ($\psi$ = 39 MPa) at $p$ = 5 MPa. It can be observed that, at high suction, heating from $T$ = 25°C to 80 °C induced an expansion and cooling from $T$ = 80 °C to 25 °C induced a contraction. The volume change during this thermal cycle is approximately reversible; a coefficient of thermal expansion $\alpha$ = $2\times10^{-4}$ °C$^{-1}$ can be estimated. At a suction of 39 MPa (test T5), as opposed to the case in test T2 ($\psi$ = 39 MPa, $p$ = 0.1 MPa), heating from $T$ = 25 °C to 80 °C resulted in a thermal contraction ($\varepsilon_v$ = 0.5%) and cooling from $T$ = 80 °C to 25 °C also resulted a contraction, leading to a total volumetric strain $\varepsilon_v \cong 1.0\%$. The volumetric strain and the temperature during cooling from $T$ = 55 °C to 25 °C can be correlated with a linear function, defining a coefficient of thermal expansion $\alpha$ = $2\times10^{-4}$ °C$^{-1}$ which is similar to that deduced from the heating phase in tests T1, T2 and T4.

Volume change under mechanical loading

The mechanical loading in tests T1, T3, T6, T7, T8, T9 and T10 were performed at constant suction and temperature. The stress paths of these tests are presented in Figure 9. Test T6 was performed at the initial total suction $\psi$ = 110 MPa and room temperature $T$ = 25 °C; Test T1 was performed at $\psi$ = 110 MPa and $T$ = 60 °C. T7, T10 and T9 were performed at the same suction $\psi$ = 39 MPa, and different temperatures: 25 °C (test T7), 40 °C (test T10), and 60 °C (test T9). T3 and T8 were performed at the same suction $\psi = 9$ MPa and different temperatures: $T = 25$ °C (test T8) and $T = 80$ °C (test T3).

Figure 10 presents the results (void ratio $e$ versus logarithm of pressure log$p$) for all these tests. From these curves, the compressibility parameters can be determined: (i) yield pressure, $p_0$; (ii) elastic compressibility parameter, $\kappa = \Delta e / \Delta \ln p$; (iii) plastic compressibility parameter, $\lambda(s) = \Delta e / \Delta \ln p$ (note that $\Delta e$ is equivalent to $\Delta v$, where $v$ is the specific volume, $v = 1 + e$). In this case, the yield pressure is determined from the intersection between the elastic compression slope and the plastic compression slope. This method is also used by Lloret et al. (2003).

For the test at $\psi = 110$ MPa, heating under p = 0.1 MPa from $T = 25$°C to 60°C increased the void ratio from 0.519 to 0.525 (T1), showing a thermal expansion as presented in Figure 7. Comparing to the volume change under mechanical loading (i.e. increased confining pressure), this thermal volume change is quite small. The two curves of tests T1 (60 °C) and T6 (25 °C) respectively are similar, showing a negligible effect of temperature (Figure 10$a$). Similar observations can be made from the three tests at 39 MPa suction (Figure 10$b$). Heating from 25 °C to 60 °C raised the void ratio from $e = 0.902$ to 0.912. The difference between the compression curves of T7 (25°C) and T9 (60°C) is not significant. The initial void ratio of the sample used in T10 (40 °C) is relatively smaller. Nevertheless, the shape of the compression curve is similar to that of T1 and T6. In test T3, the thermal cycle, $T = 25 - 80 - 25 - 80$ °C, reduced the volume of the soil specimen at $\psi = 9$ MPa due to thermal contraction (Figure 7). In Figure 10$c$, it can be observed that this contraction corresponds to a decrease of the void ratio from 1.290 to 1.267. The comparison between the compression curves from T8 (25 °C) and T3 (80 °C) does not show any significant temperature effect: the two curves are similar.

It can be observed from Figure 10 that $\lambda(s)$ and $\kappa$ are independent of temperature while they are strongly affected by suction. The values of these parameters are plotted versus suction in Figure 11 (only for the cases of tests at 25 °C). It appears that wetting (decrease of total suction) increased the compressibility parameters.

The yield pressures ($p_0$) determined from Figure 10 are shown in Figure 12 for different suctions and temperatures. It can be observed that $p_0$ decreased from 17.1 MPa at the initial suction $\psi$ = 110 MPa to 2.1 MPa at $\psi$ = 39 MPa and 0.38 MPa at $\psi$ = 9 MPa for tests at 25 °C. At $\psi$ = 110 MPa, heating from $T$ = 25 °C to 60 °C decreased $p_0$ from 17.1 MPa (T6) to 12.8 MPa (T1). At $\psi$ = 39 MPa, heating decreased equally $p_0$ from 2.1 MPa (T7, 25°C) to 1.6 MPa (T10, 40 °C) and 0.8 MPa (T9, 60 °C). And at $\psi$ = 9 MPa, $p_0$ was decreased from 0.38 MPa (T8, 25 °C) to 0.32 MPa (T3, 80 °C).

Figure 13 presents the stress paths of two tests (T5 and T11) where a mechanical loading was applied after a thermal cycle at high pressure. For test T11, from the initial conditions ($\psi$ = 110 MPa, $p$ = 0.1 MPa, $T$ = 25 °C), the soil sample was first loaded to $p$ = 20 MPa, then subjected to a thermal cycle, $T$ = 25 – 80 – 25 °C, under constant pressure $p$ = 20 MPa. Afterwards, at $T$ = 25 °C, the specimen was compressed up to $p$ = 50 MPa. For test T5, the soil specimen was first wetted to $\psi$ = 39 MPa and then loaded to $p$ = 5 MPa. Under this pressure, a thermal cycle, $T$ = 25 – 80 – 25 °C, was applied, followed by loading up to $p$ = 20 MPa.

The corresponding void ratio changes are presented in Figure 14. The initial loading from $p$ = 0.1 MPa to 20 MPa in T11 decreased the void ratio from 0.513 to 0.455. After the thermal cycle under $p$ = 20 MPa, the void ratio was reduced to 0.448, showing that a plastic thermal contraction occurred. Loading from $p$ = 20 MPa to 50 MPa showed an elasto-plastic behaviour with a yield pressure, $p_0$ = 36.1 MPa. A similar behaviour can be observed on test T5. The thermal cycle at $\psi$ = 39 MPa and $p$ = 5 MPa reduced the void ratio from 0.774 to 0.754 (this corresponds to the thermal contraction observed in Figure 8). An elasto-plastic behaviour can also be observed during the compression from $p$ = 5 MPa to 20 MPa. The yield pressure $p_0$ was estimated to be 7.9 MPa, which is clearly higher than the constant pressure during the thermal cycle (5 MPa).

**Discussion**

In the present work, mechanical loading (i.e. increasing the confining pressure) was performed step-by-step to allow suction equilibrium after each loading step. A disadvantage of this approach is that the evaluation of compressibility parameters can be somewhat approximate when there are not enough data. In the thermo-mechanical study of saturated soils, a constant stress rate loading is usually used (Sultan *et al.* 2002) which allow a more accurate determination of the compressibility parameters. However, constant stress rate loading can not be used when unsaturated soil is concerned as suction control can not be checked. For this reason, step-by-step mechanical loading is usually used to study the hydro-mechanical behaviour of unsaturated soil (Lloret *et al.* 2003; Romero *et al.* 2005). In the present work, in spite of the approximate evaluation of yield pressure and compressibility parameters, several effects of suction and temperature can be observed.

It was observed that the volumetric thermal behaviour is strongly affected by suction and pressure. Heating induced expansion under low pressure and high suction; at high pressure and low suction, heating tended to induce a contraction (Figures 7 and 8). In the case of saturated soils, Plum & Esrig (1969), Baldi *et al.* (1988), Towhata *et al.* (1993), Del Olmo *et al.* (1996), Robinet *et al.* (1997), Burghignoli *et al.* (2000), Sultan *et al.* (2002), Cekerevac *et al.* (2003) observed that heating a clay at low overconsolidation ratio (*OCR*) induced plastic contraction. These authors observed equally that, at high *OCR*, heating induced expansion up to a certain temperature; a contraction took place when the temperature was higher. These observations are in good agreement with the results obtained on unsaturated MX80 bentonite in the present work. Indeed, the results obtained on $p_0$ presented in Figure 12 can be used to calculate *OCR*. For T1 and T2 (Figure 7) where thermal expansion occurred during heating, *OCR* values were large: for T1, $\psi = 110$ MPa, $OCR = \dfrac{p_0}{p} = \dfrac{17.1}{0.1} = 171$; for T2, $\psi = 39$ MPa, $OCR = \dfrac{p_0}{p} = \dfrac{2.1}{0.1} = 21$. For T5, loading until $p = 5$ MPa exceeded the initial $p_0$ ($p_0 = 2.1$ MPa at $\psi = 39$ MPa). The soil was thus heated in a normally consolidated state (*OCR* = 1). It can be concluded that heating induced expansion at high *OCR* and contraction at *OCR* = 1 in the case of unsaturated soils. It can be also concluded that *OCR* is not the only parameter that governs the thermal volumetric behaviour of an unsaturated soil. In fact, thermal contraction occurred in T3 during heating (*OCR* = 3.8, $\psi = 9$ MPa, $p = 0.1$ MPa). On the contrary, at a similar *OCR* value (*OCR* = 3.4) heating induced expansion in T4 ($\psi = 110$ MPa, $p = 5$ MPa). The suction effect is thus clearly evidenced, which can be explained by the softening of the swelling clay aggregates due to suction decrease.

The works of Del Olmo *et al.* (1996), Robinet *et al.* (1997), Sultan *et al.* (2002) showed that during thermal cycles on saturated clays: (i) the contraction during heating is irreversible and the contraction during cooling is reversible; (ii) the expansion during heating is reversible; (iii) the slopes $d\varepsilon_{vt}/dT$ (change in thermal volumetric strain / change in temperature) of the expansion curve during heating and the contraction curve during cooling are similar; this value corresponds to the soil's coefficient of thermal expansion. These observations are in good agreement with the results obtained in the present work (Figures 7 and 8). Indeed, the heating/cooling cycle ($T$ = 25 – 80 – 25 °C) induced an irreversible volumetric strain $\varepsilon_{vt}$=1.0% in T3 and T5. The re-heating that followed the cooling in T3 showed that the contraction during cooling is reversible. In addition, the cooling phase following the heating phase in T4 equally evidenced that the expansion during heating is reversible. The coefficients of thermal expansion determined from tests T1, T2, T4, and T5 were similar, $\alpha = d\varepsilon_v/dT = 2\times10^{-4}(°C^{-1})$ ; for T3, the data scatter has not enabled this determination. This value is of the same order of magnitude of that measured by Romero *et al.* (2005) on an unsaturated compacted expansive clay at low confining pressures.

As far as the suction effect on the compressibility parameters ($\lambda(s)$ and $\kappa$) is concerned, (Figure 11), the soil volume change at low stresses is mainly governed by the compressibility of soil aggregates. In other words, the elastic compressibility parameter ($\kappa$) mainly depends on the compressibility of soil aggregates. In the case of low plasticity soils, it has been observed that $\kappa$ is slightly dependent on the suction (Cui & Delage, 1996; Cuisinier & Masrouri, 2004; Alshihabi *et al.,* 2002). On the contrary, this parameter increases with a decrease of suction in the case of expansive clays (Al-Mukhtar *et al.,* 1999; Lloret *et al.,* 2003; Marcial, 2003). In fact, upon wetting, the expansive clay aggregates swell, giving rise to a mechanical softening. As a result, the compressibility of the aggregates is increased, thus increasing the elastic compressibility parameter ($\kappa$). It is not the case with low plasticity soils where no significant expansion occurs. During compression at high pressures, the soil volume change is the sum of the volumetric strain of soil aggregates and the collapse of inter-aggregate macro-pores. The plastic compressibility parameter ($\lambda(s)$) is then strongly affected by the suction whatever the soil nature. Al-Mukhtar *et al.* (1999) observed that the effect of suction on $\lambda(s)$ is more significant for higher plasticity soils. Alshihabi *et al.* (2002) observed that this parameter decreases slightly after a wetting/drying cycle. Cuisinier & Masrouri (2005) concluded that $\lambda(s)$ depends strongly on the soil's hydro-mechanical history. Regarding the yield pressure $p_0$, its decrease due to wetting (Figure 12) is a well-known phenomenon in the mechanics of unsaturated soils.

The temperature effect on the compressibility of unsaturated soil was not significant in the present work. The results presented in Figure 10 showed that $p_0$ is slightly influenced by temperature but the compressibility parameters ($\lambda(s)$ and $\kappa$) are insensitive to the temperature changes. Work on saturated clays has equally shown the independence of these compressibility parameters on temperature (Campanella & Mitchell, 1968; Fleureau, 1972; Habibagahi, 1977; Belanteur *et al.*, 1997; Burghinoli *et al.*, 2000; Cekerevac & Laloui, 2004). Only Sultan *et al.* (2002) observed a decrease of $\lambda$ after cooling from $T = 100$ °C to 20 °C on Boom clay. For unsaturated soils, Recordon (1993) and Saix *et al.* (2000) observed the temperature independence of $\lambda(s)$ and $\kappa$ for sand and silty clayey sand. On the contrary, Romero *et al.* (2003) observed a temperature insensitivity of $\kappa$ but a temperature dependence of $\lambda(s)$ for compacted Boom clay. It appears that the temperature effect depends on the thermal expansion of the soil aggregates prior to mechanical compression. If this expansion is not significant, the soil aggregates are not significantly weakened, and in that case, the compressibility parameters must be temperature-independent. On the contrary, when significant thermal expansion takes place, a temperature-dependent compressibility behaviour is expected.

The results presented in Figure 12 evidenced the temperature effect on $p_0$. In all cases, heating under $p = 0.1$ MPa decreased $p_0$. This temperature effect on $p_0$ was equally observed on saturated soils by Baldi *et al.* (1991), Towhata *et al.* (1993), Sultan *et al.* (2002) and Laloui & Cekerevac (2003).

The thermal hardening phenomenon can be observed in Figure 14. A thermal cycle was performed at a constant pressure that was higher than the initial yield pressure. In the case of T11 ($\psi = 110$ MPa, $p_0 = 17.1$ MPa in Figure 12), the soil was normally consolidated at $p = 20$ MPa. However, after a thermal cycle under this pressure, an overconsolidated behaviour was observed with a new $p_0$ value of 36.1 MPa. The same phenomenon was observed in T5 ($\psi = 39$ MPa, $p_0 = 2.1$ MPa in Figure 12). The thermal cycle at a constant pressure of 5 MPa resulted in an increase of $p_0$ to 7.9 MPa. In fact, the thermal cycle induced contraction which hardened the soil and therefore increased $p_0$. This thermal hardening phenomenon was equally observed by Sultan *et al.* (2002) on saturated Boom clay.

**Conclusions**

Thermo-mechanical tests were performed on unsaturated heavily compacted MX80 bentonite using a suction-temperature controlled isotropic cell. The results from thermal loading tests at constant pressure showed the effects of suction and pressure on the volumetric thermal behaviour of soil. Heating induced expansion under low pressure and high suction. At high pressure and low suction, heating tended to induce a contraction. For unsaturated soils, it appears that the over-consolidation ratio and the suction are both important parameters that govern the thermal volumetric behaviour. The tests with heating/cooling cycles evidenced that the thermal expansion is reversible. If thermal contraction occurs during heating, the yield pressure may be increased due to a thermal hardening phenomenon.

The results from mechanical loading tests evidenced the effects of suction and temperature on the compressibility behaviour. It appears that a reduction in suction gives rise to increased compressibility parameters and a decrease of yield pressure. However, the temperature effect on the compressibility parameters ($\kappa$ and $\lambda(s)$) was not significant. In addition, heating induced a slight decrease in the yield pressure.

## Acknowledgement

The authors are grateful to Ecole Nationale des Ponts et Chaussées (ENPC) and French Electricity Company (EDF) for their financial support.

**Table 1. Stress paths of tests. (All the tests start with the same initial conditions: p = 0.1 MPa; $\psi$ = 110 MPa; T = 25 °C).**

| Test No. | Path | | | | | | | | | | | | | | |
|---|---|---|---|---|---|---|---|---|---|---|---|---|---|---|---|
| | I | | | II | | | III | | | IV | | | V | | |
| | p: MPa | $\psi$: MPa | T: °C | p: MPa | $\psi$: MPa | T: °C | p: MPa | $\psi$: MPa | T: °C | p: MPa | $\psi$: MPa | T: °C | p: MPa | $\psi$: MPa | T: °C |
| T1 | 0.1 | 110 | 60 | 50 | 110 | 60 | | | | | | | | | |
| T2 | 0.1 | 39 | 25 | 0.1 | 39 | 70 | | | | | | | | | |
| T3 | 0.1 | 9 | 25 | 0.1 | 9 | 80 | 0.1 | 9 | 25 | 0.1 | 9 | 80 | 5 | 9 | 80 |
| T4 | 5 | 110 | 25 | 5 | 110 | 80 | 5 | 110 | 25 | | | | | | |
| T5 | 0.1 | 39 | 25 | 5 | 39 | 25 | 5 | 39 | 80 | 5 | 39 | 25 | 20 | 39 | 25 |
| T6 | 60 | 110 | 25 | | | | | | | | | | | | |
| T7 | 0.1 | 39 | 25 | 10 | 39 | 25 | | | | | | | | | |
| T8 | 0.1 | 9 | 25 | 5 | 9 | 25 | | | | | | | | | |
| T9 | 0.1 | 39 | 25 | 0.1 | 39 | 60 | 50 | 39 | 60 | | | | | | |
| T10 | 0.1 | 39 | 25 | 0.1 | 39 | 40 | 50 | 39 | 40 | | | | | | |
| T11 | 20 | 110 | 25 | 20 | 110 | 80 | 20 | 110 | 25 | 50 | 110 | 25 | | | |

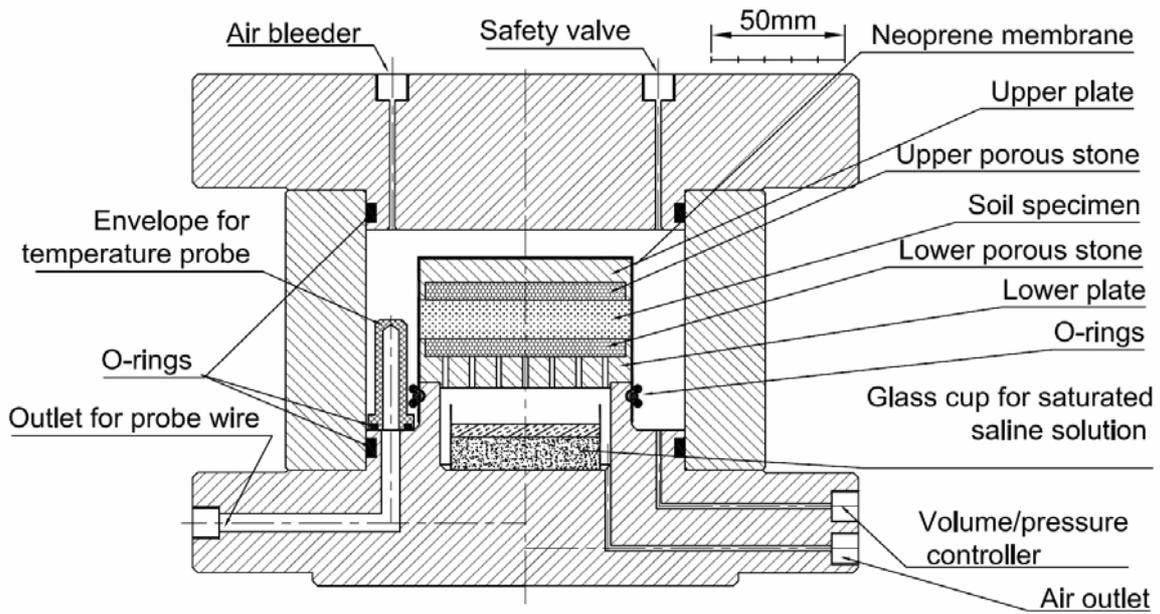

**Figure 1. Schematic view of the isotropic cell**

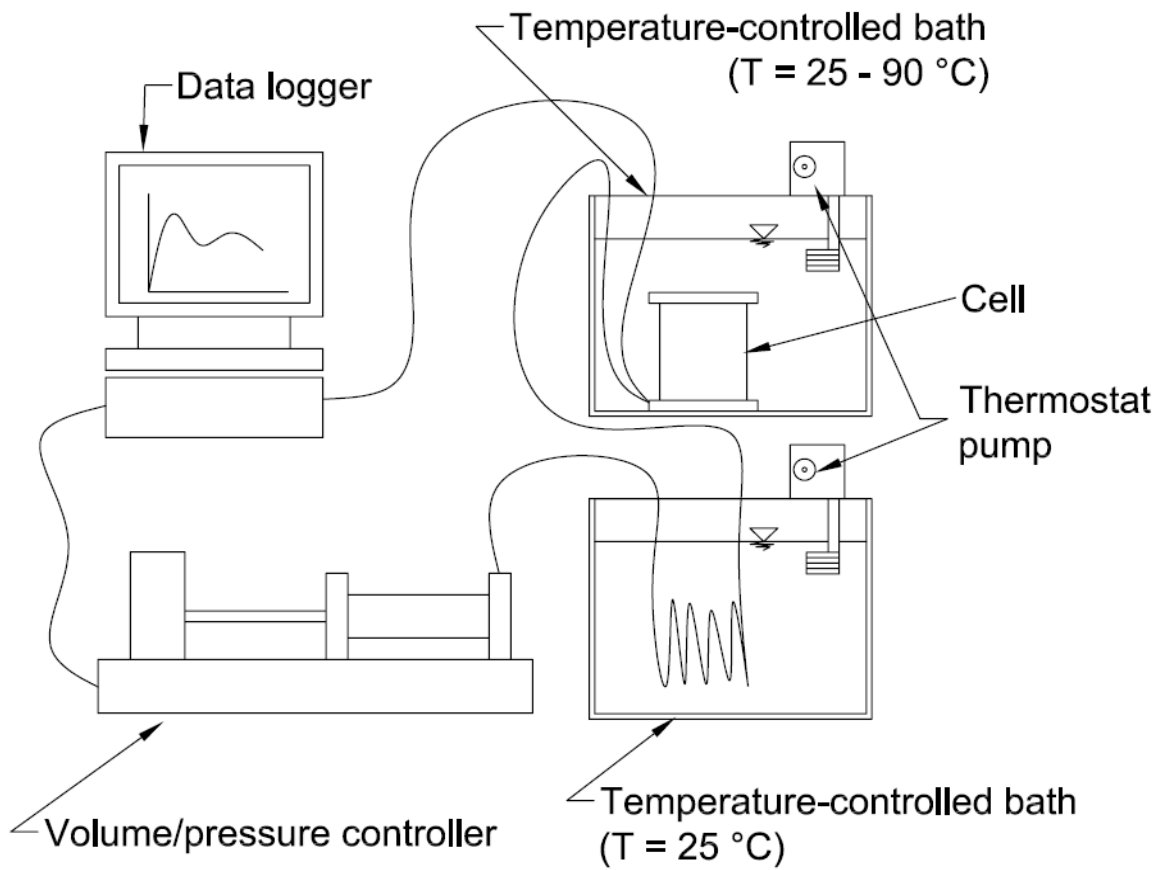

**Figure 2. Experimental setup of suction-temperature controlled isotropic test.**

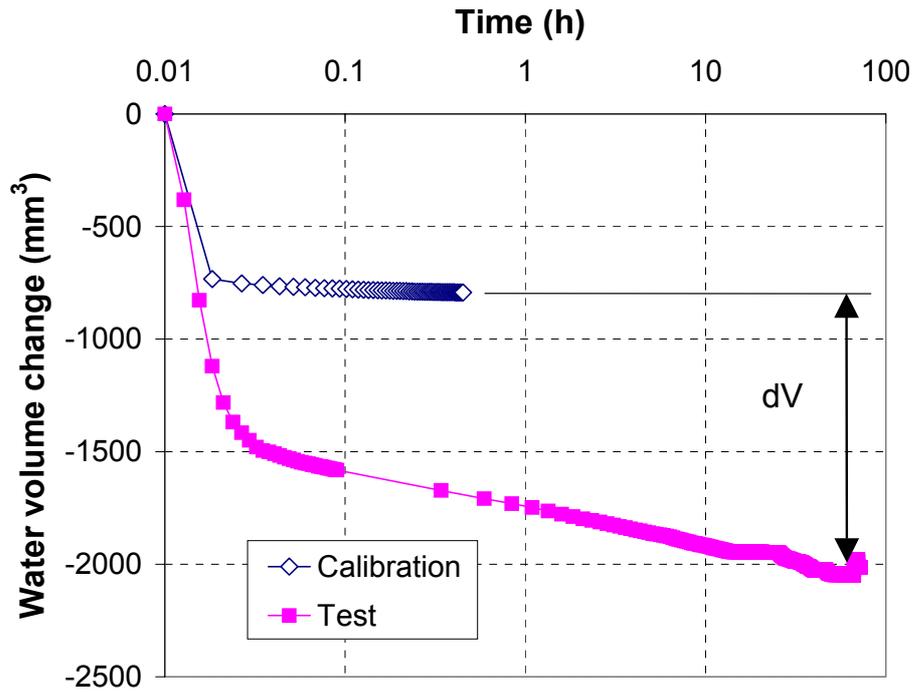

**Figure 3.** Determination of soil volume change during a loading step from 1 to 2 MPa.

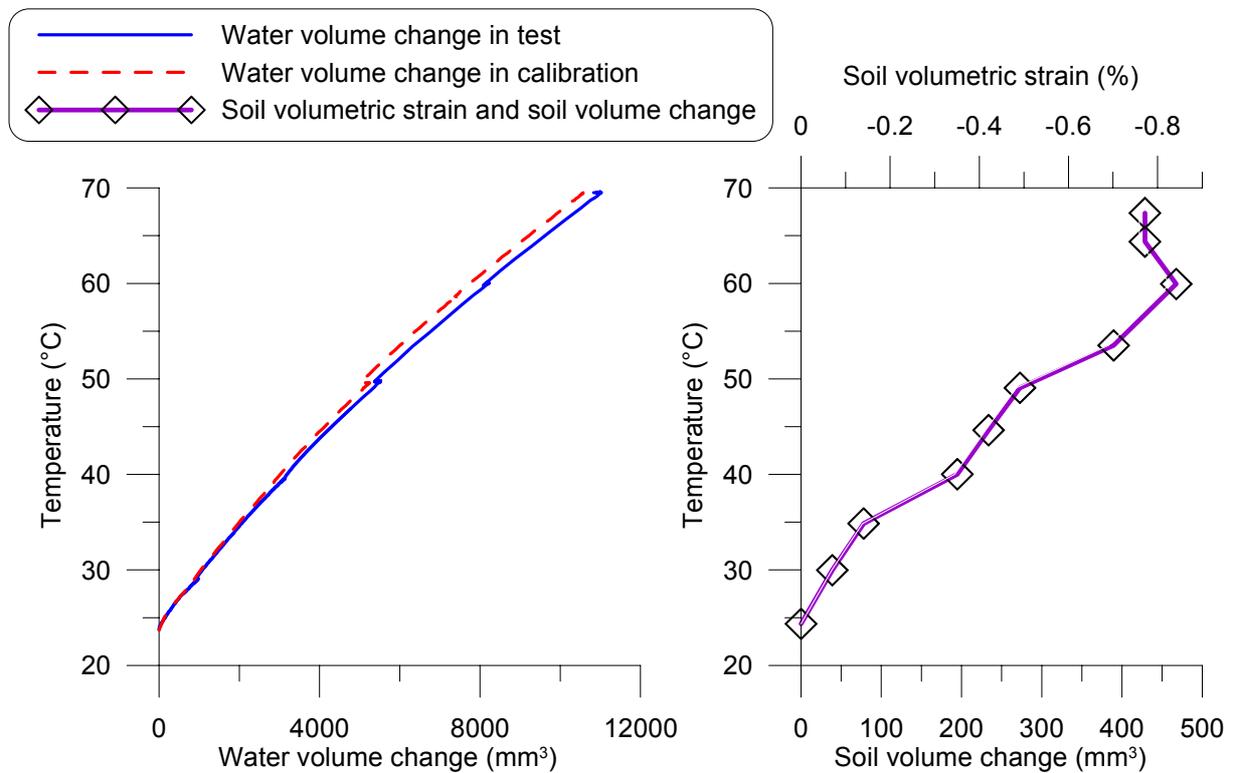

**Figure 4.** Determination of soil volume change during heating under confining pressure of 0.1 MPa.

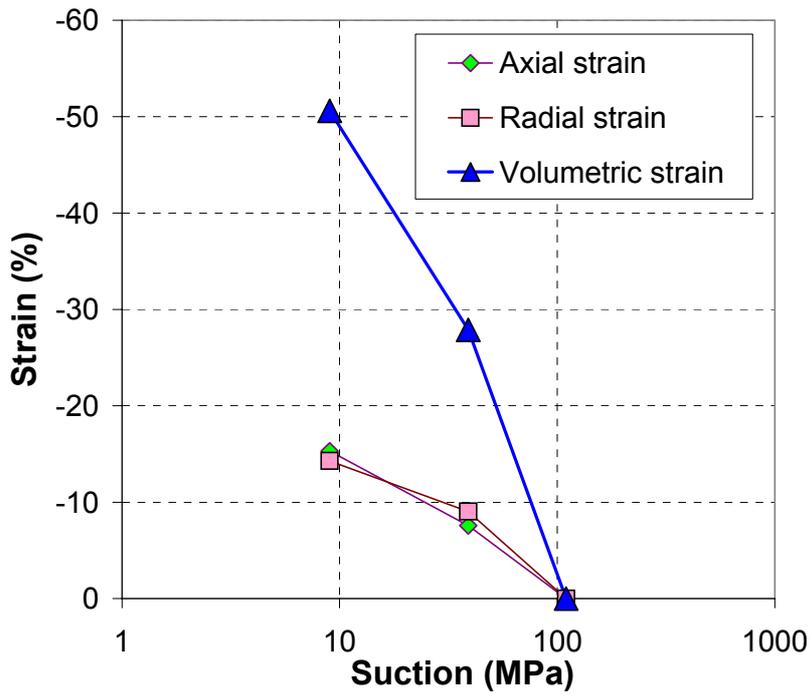

**Figure 5. Axial, radial, and volumetric strain during wetting.**

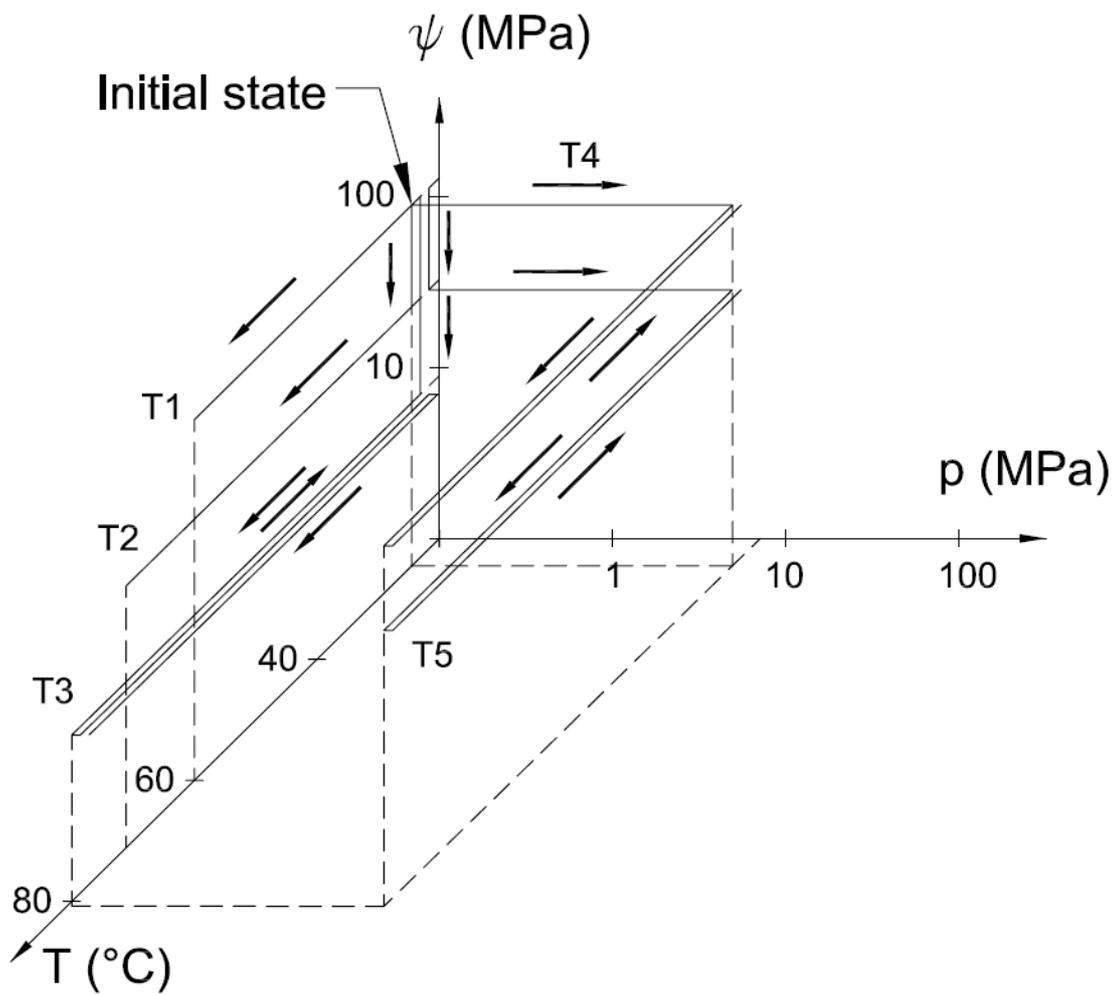

**Figure 6. Stress paths of tests T1, T2, T3, T4, T5 (for studying the thermal volumetric behaviour).**

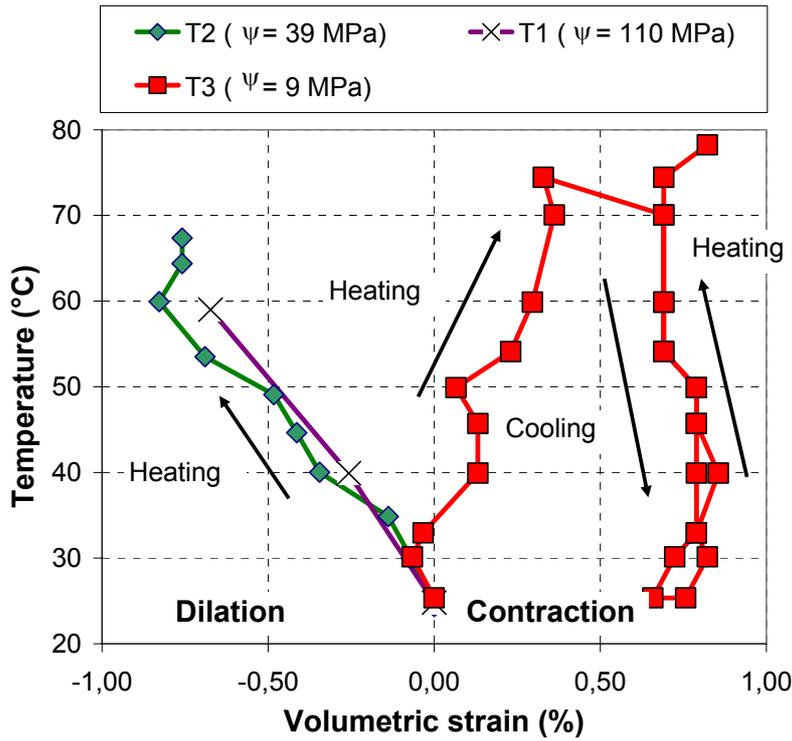

**Figure 7. Volumetric strain during thermal loading under constant pressure at 0.1 MPa. Tests T1, T2, T3.**

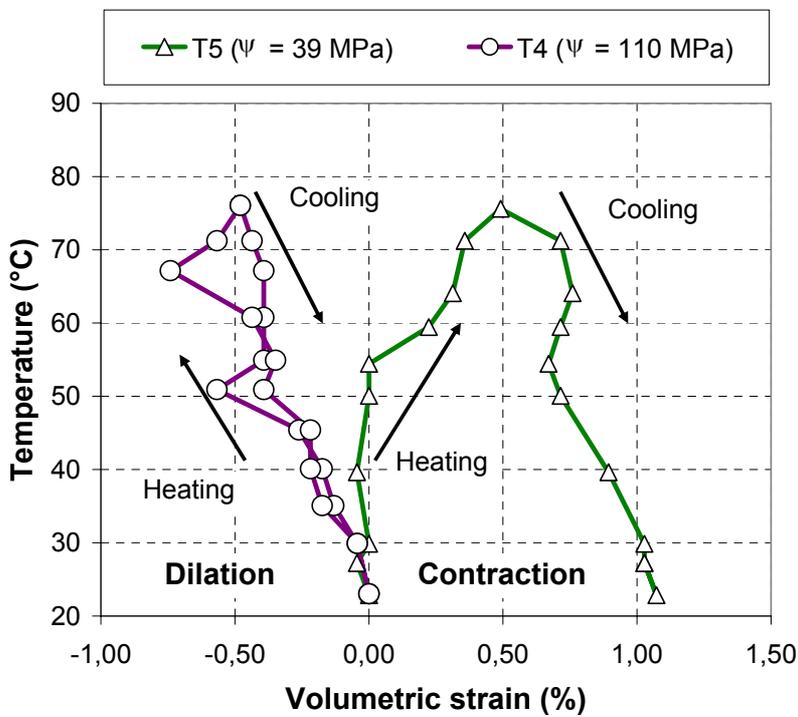

**Figure 8. Volumetric strain during thermal loading under constant pressure at 5 MPa. Tests T4, T5.**

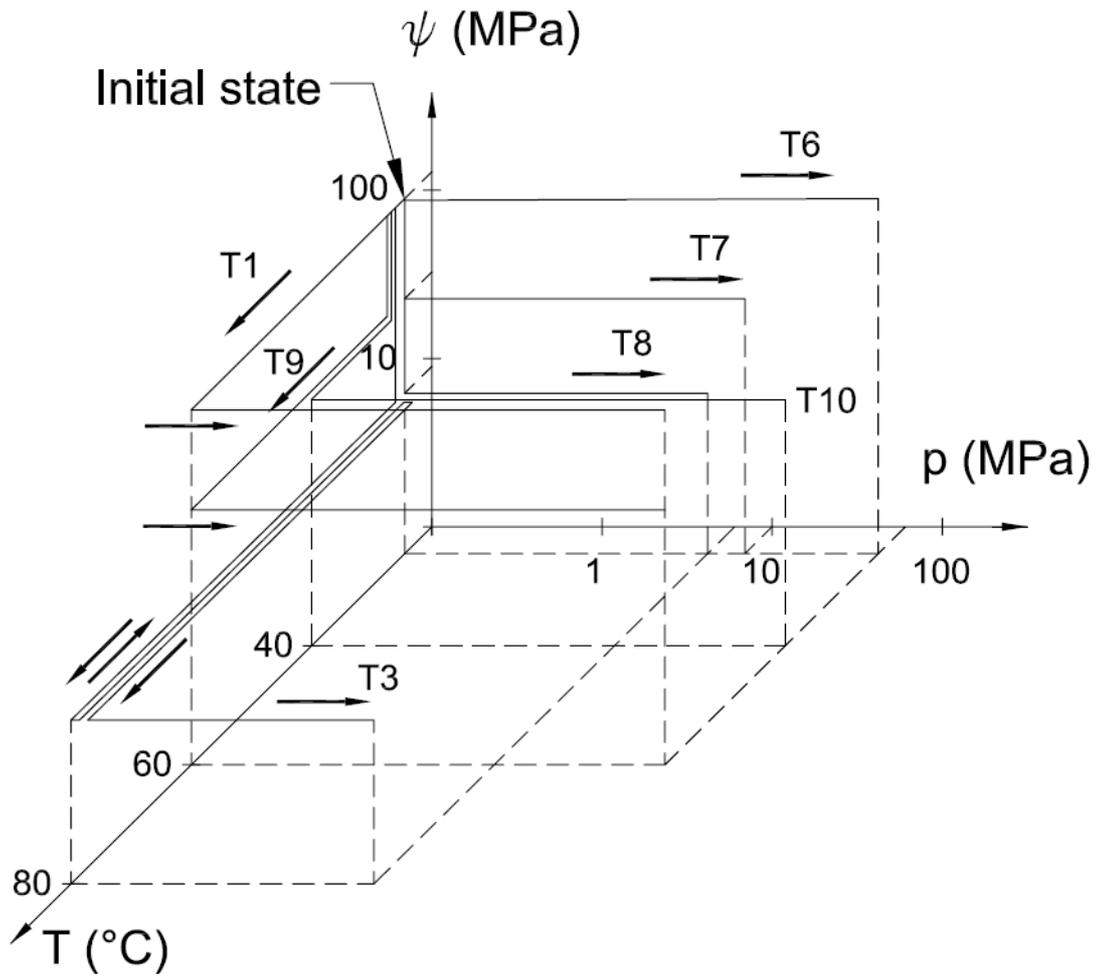

**Figure 9.** Stress paths of tests T1, T3, T6, T7, T8, T9, and T10 (for studying the mechanical volumetric behaviour).

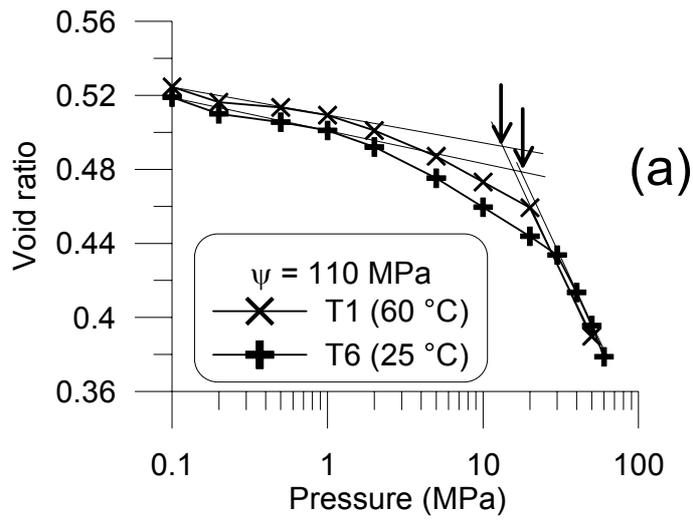
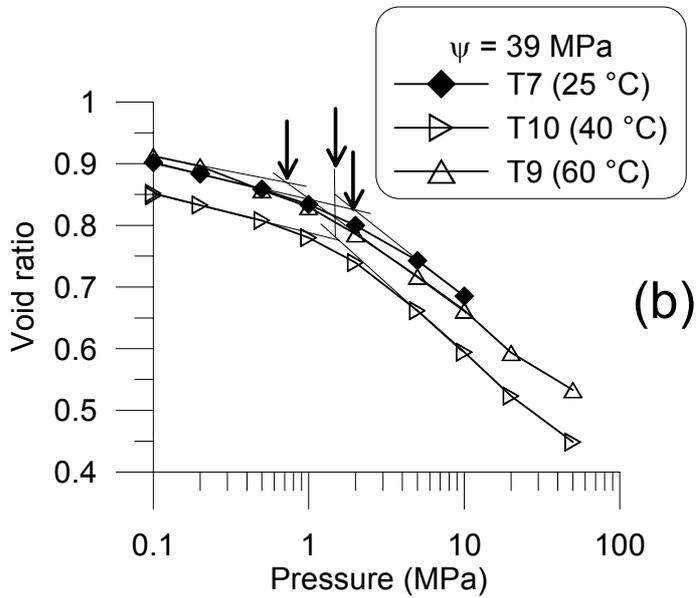
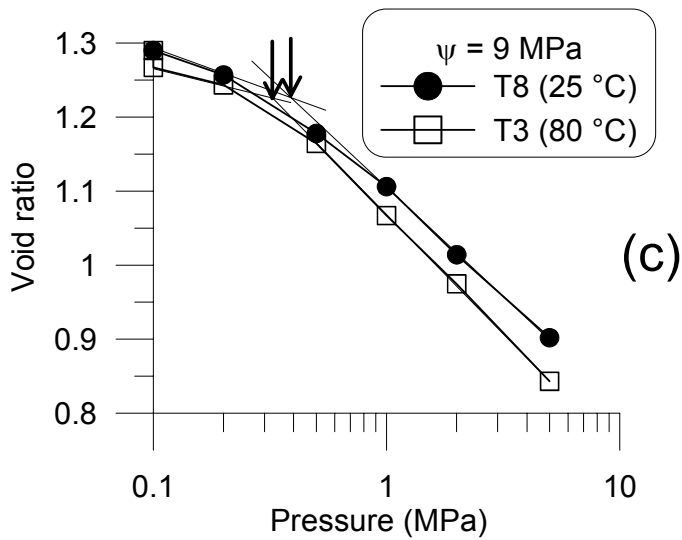

**Figure 10. Results obtained from mechanical loading at constant suction and temperature. Tests T1, T3, T6, T7, T8, T9, and T10.**

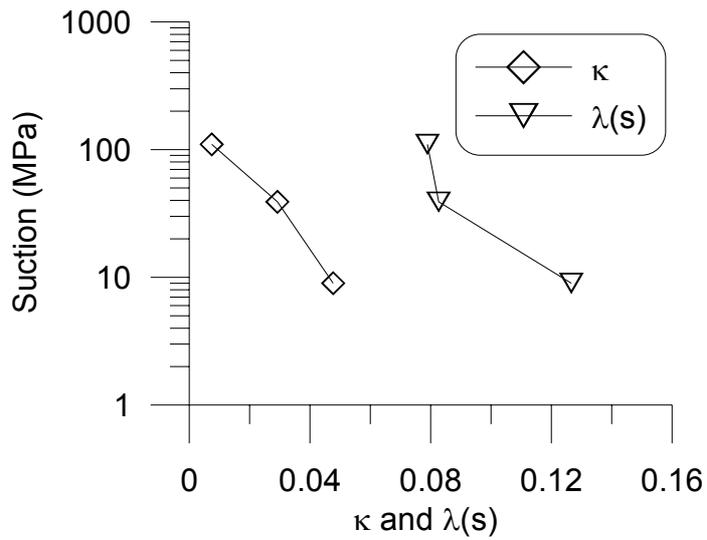

**Figure 11. Compressibility parameters ($\kappa$ and $\lambda(s)$) versus suction for tests at 25 °C.**

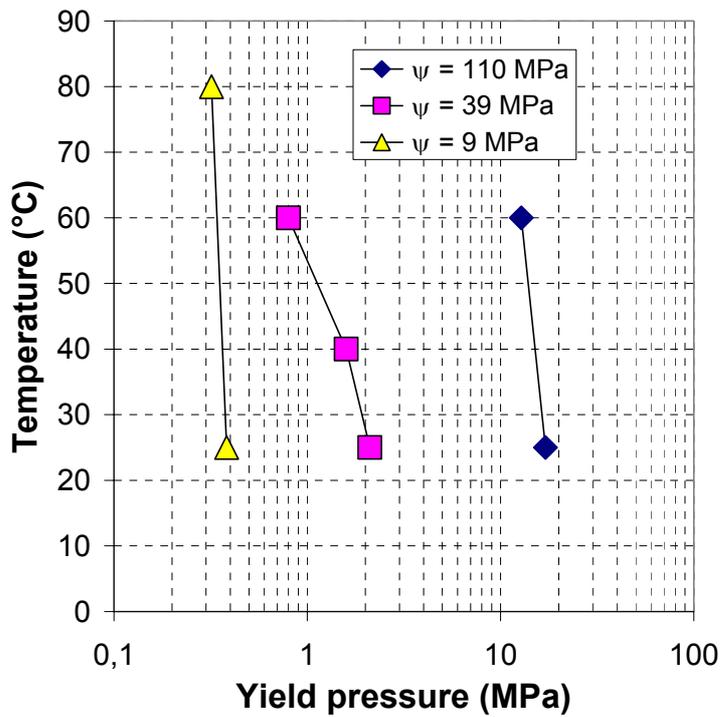

**Figure 12. Yield pressure ($p_0$) versus temperature for different suctions.**

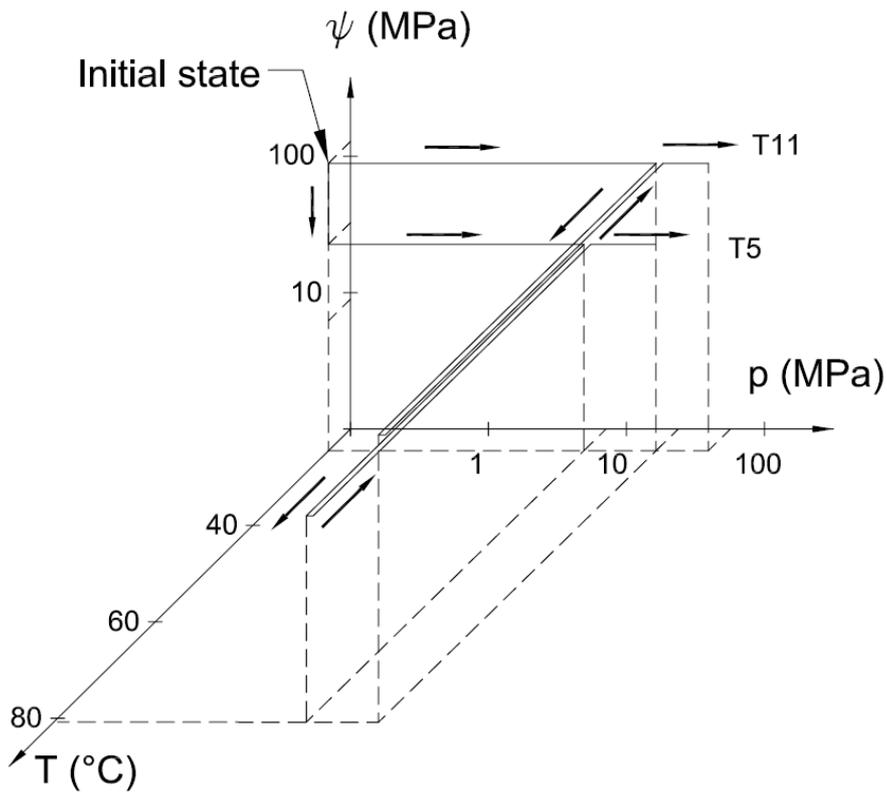

Figure 13. Stress paths of tests T5 and T11 (for studying the thermal hardening).

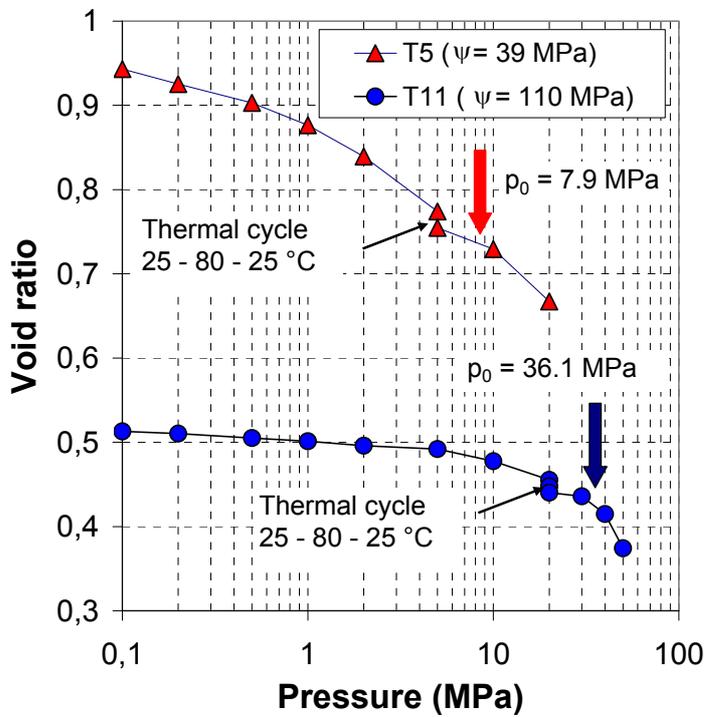

Figure 14. Void ratio change under thermo-mechanical loading for different suction. Tests T5 and T11.